\documentstyle[12pt]{article}
\newtheorem{theorem}{Theorem}

\title{On the algebraic structures connected with the
       linear Poisson brackets of hydrodynamics type}

\author{A A Balinsky\thanks{\mbox{Permanent address:}
Technion-Israel Institute of Technology, Department of Mathematics, \ \ \ \ \ \
32000 Haifa, ISRAEL}
 \ and   A I Balinsky \\
 \small Institute for Applied Problems
 \small of Mechanics and Mathematics \\
 \small     Academy of Sciences of the Ukraine \\
 \small    $ 3^{a} $ Naukova Str., 290601 Lvov, Ukraine }
\date{}

\begin{document}

\maketitle

\begin{abstract}  \small
     The generalized form of
     the Kac formula
     for Verma modules
     associated with linear
     brackets of hydrodynamics
     type is proposed.
     Second cohomology groups
     of the generalized
     Virasoro algebras are
     calculated. Connection
     of the central extensions
     with the problem of
     quntization of hydrodynamics
     brackets is demonstrated.
\end{abstract}

{\bf 1.} Poisson brackets of hydrodynamics type (PBHT)                
\begin{equation}
           \{u^{i}(x),u^{j}(y)\}=g^{ij}(u(x))\delta'(x-y)+
           u^{k}_{x}b^{ij}_{k}(u(x))\delta(x-y)            \label{hyd}
\end{equation}
(here and below we assumed a summation on repeat
indexes) were introduced and studied in \cite{DN1,DN2}
to construct a theory of conservative systems of
hydrodynamics type and to develop a Bogolubov-Whitham
method of averaging Hamiltonian field-theoretic systems.
We refer to the recent expository article~\cite{DN3}
and the extensive bibliography therein.
In \cite{BalNov} S~P~Novikov and first author
considered and gave a classification of  these Poisson
brackets depending linearly on the fields $u^{j}$
relative to linear change $u^{k}=A^{k}_{j}w^{j}$.
Some examples were discussed in \cite{Nov,GelDor}.

For the reader's convenience we recall some construction
from \cite{BalNov}. The simplest local Lie algebras arising
from the brackets of hydrodynamics type are especially
interesting, where, according to \cite{BalNov},
in the case when all metrics are linear in $u$ we have
\begin{eqnarray}
     g^{ij} &  = & C^{ij}_{k}u^{k}+g^{ij}_{0}     \nonumber  \\
     b^{ij}_{k} & = & const, \ \ g^{ij}_{0}=const  \\  \label{lin}
     C^{ij}_{k} & = & b^{ij}_{k}+b^{ji}_{k}.       \nonumber
\end{eqnarray}
 Linear (\ homogeneous\ ) part of such PBHT determines
 some
 very interesting class of the infinite-dimensional
 Lie algebras (``hydrodynamic algebras''):
 for two vector-functions $f(x)$ and $g(x)$ with $N$
 components $f_{p}$, $g_{q}$ we may define the commutator
 in the ``local translation-invariant
 first-order Lie algebra'' or the hydrodynamic algebra
\begin{equation}
 [ f,g ]_{k}(x)  =
 b^{pq}_{k}(f_{p}^{'}(x)g_{q}(x)-g_{p}^{'}(x)f_{q}(x)).  \label{alg}
\end{equation}
A bracket (\ref{hyd}) or Lie algebra (\ref{alg}),
linear in the fields,
is called {\em symmetric\/} if  $b^{ij}_{k}=b^{ji}_{k}$.
Here $f_{p}$ are adjoint variables for $u^{i}$.

It is useful to introduce a new algebra {\bf B} as a
multiplication in $N$-space
$M$ with basis $e^{1},e^{2},\ldots,e^{N}$
\begin{equation}
       e^{i}e^{j}=b^{ij}_{k}e^{k}. \label{myalg1}
\end{equation}
For the functions $f(x)=f_{p}(x)e^{p}$ and
$g(x)=g_{q}(x)e^{q}$
we write (\ref{alg}) in the form $f^{'}g-g^{'}f$ using
multiplication (\ref{myalg1}) in the algebra {\bf B}.
The tensor \(b^{ij}_{k} \) defines by (\ref{alg}) a local
translationally invariant Lie algebra of first order if
and only if the multiplication law (\ref{myalg1}) defines
a finite-dimensional algebra {\bf B} in which the following
identities hold:
\begin{equation}
       a, b, c \in {\bf B}, \ \ \ (ab)c=(ac)b,
       \ \ \ \ (ab)c-a(bc)=(ba)c-b(ac) \label{myalg2}.
\end{equation}
In the symmetric case
$2b^{ij}_{k}=2b^{ji}_{k}=C^{ij}_{k}$
this algebra is
commutative and associative.
                                                             \\

{\bf Remark} \ \ \ After introducing an operation                     
$[a,b]=ab-ba$
on {\bf B} it is Lie algebra and as was first proved in
 \cite{BalSup} this is a solvable Lie algebra.
If a finite-dimensional algebra {\bf B}
(\ref{myalg1}),(\ref{myalg2})
is commutative then it is automatically associative,
and if it has the right unit then it's commutative.
The theory of extensions for the algebras
 {\bf B} (\ref{myalg1}),(\ref{myalg2})
was constructed in \cite{BalNov,Zel}. The method
of construction the wide class of such algebras was
proposed by S~I~Gelfand:
if algebra $A$ is commutative and associative,
$\partial$ - differentiation of $A$
then multiplication $a\circ b=a(\partial b)$
satisfied (\ref{myalg2}).
                                                                     \\

 The formula (\ref{hyd}) define Poisson bracket if and only if an
 algebra (\ref{myalg1}) satisfied (\ref{myalg2}) and for the
 following symmetric belinear product
         \[( e^{i},e^{j})_{0}=g^{ij}_{0}\]
we have
\begin{equation}
      (ab,c)_{0}=(a,cb)_{0} \label{fr},
\end{equation}
for all $a,b,c \in \mbox{\bf B}$. In this case the
`` quasifrobenius property '' (\ref{fr})
hold true for all symmetric belinear products with matrix
$g^{ij}(u)   =  C^{ij}_{k}u^{k}+g^{ij}_{0}$ for any $u^{i}$.
If the algebra {\bf B} is commutative and has the unit then we
have the classical {\em frobenius algebra\/}.

Poisson bracket (\ref{hyd}) is
called {\em nondegenerate\/} if the
pseudo-Riemannian metric
$g^{ij}  =  C^{ij}_{k}u^{k}+g^{ij}_{0}$
is nondegenerate at a "generic point":
  \[ det( g^{ij}):=P_{N}(u^{1},u^{2},...u^{N}) \not\equiv 0.\]
Lie algebra (\ref{alg}) and a finite-dimensional algebra
{\bf B} (\ref{myalg2})
are called {\em nondegenerate} if the pseudo-Riemannian metric
$ g^{ij}  =  C^{ij}_{k}u^{k} $ is nondegenerate
at a "generic point",
where $b^{ij}_{k}+b^{ji}_{k}=C^{ij}_{k}.$ From the main
theorem of \cite{DN1} we know that in a nondegenerate case
these pseudo-Riemannian metrics have a vanishing curvature.

{\bf 2.} \ \ \ For  the vector--valued functions periodic in $x$,      
by passing to an expansion in Fourier series, we obtain a basis
$(L^{i}_{n})$ for the algebra (\ref{alg}) with the relations
\begin{equation}
[L^{i}_{n},L^{j}_{m}]=(nb^{ij}_{k}-mb^{ji}_{k})L^{k}_{m+n}, \label{W}
\end{equation}
which we  call the generalized Witt algebra.
For this algebra we have the following generalization
of the Gelfand--Fuks theorem \cite{GF} on the central
extension of the algebra of vector fields on the circle:
                                                                    %
                                                                    %
\begin{theorem}
If the algebra \mbox{\rm (\ref{myalg1})} is commutative
and has the unit then
       \[ H^{2}(V)=\bf B^{\ast},\]
where $V$ -- algebra \mbox{\rm (\ref{W})} and
$H^{2}$--second cohomology
group of Lie algebra and $\bf B^{\ast}$
is the dual space for the algebra \mbox{\rm (\ref{myalg1})}.
The all   central extensions of
\mbox{\rm (\ref{W})} have the following form
  \begin{equation}
   [L^{i}_{n},L^{j}_{m}]^{'}=(n-m)b^{ij}_{k}L^{k}_{m+n}+b^{ij}_{k}
   l^{k}\frac{(n^{3}-n)}{12}\delta_{n+m,0}Z ,           \label{vir}
  \end{equation}
where Z -- central element and $l=(l_{i}) \in \bf B^{\ast}$.
\end{theorem}
In the case when {\bf B=C} ( algebra of complex numbers )
we have Gelfand--Fuks theorem.

For the algebra (\ref{vir}) we may consider Verma module $V_{h,l}$,
\ \ \ $h,l \in \bf B^{\ast}$, over this algebra: \\
$V_{h,l}$  free generate by $\mid v>$ over $L^{i}_{n}$
with $n>0$, $i=1, \ldots ,N$ and
\begin{eqnarray}
L^{i}_{n} \mid v> & = & 0,\ \ n<0;            \nonumber     \\
L^{i}_{0} \mid v> & = & h(e^{i}) \mid v> ;       \nonumber  \\
Z \mid v> & = & \mid v >.  \nonumber
\end{eqnarray}
An element of the Verma module is singular if
it is generates the Verma submodule, i.e. it
is eigenvector for all $L^{i}_{0}$, and
annihilated by the all $L^{i}_{n}$, $n<0$.
                                                                     %
                                                                     %
\begin{theorem}
      For the algebra \mbox{\rm (\ref{vir})} with
      the unital $\bf B$ Verma module is reducible
      if and only if it has the singular vector.
\end{theorem}
When algebra (\ref{myalg1}) is commutative and
has the unit then for (\ref{vir}) from the
root decomposition of {\bf B}  we  have the
following analog of the famous
Kac--Feigin--Fuks criteria:
                                                                       %
                                                                       %
\begin{theorem}
       Verma module $V_{h,l}$ is reducible if
       and only if in the algebra {\bf B}
      \mbox{\rm (\ref{myalg1})} exist one-dimensional
       ideal $<a>$ such that if $a^{2}=0$
       then $h(a)=l(a)(\alpha ^{2}-1)/2$
       for some $\alpha \in N_{+}$, or
       if $aa=\mu a$ with $\mu \in \protect\mbox{\bf C},
       \mu \neq 0$
       then $\overline{h}:=h(a)/\mu $,
       $\overline{c}:=l(a)/\mu $
       satisfied to Kac condition
       \mbox{\rm \cite{Kac}} for
       the ordinary Virasoro algebra.
\end{theorem}

{\bf 3.} \ \ \ In this part we consider the problem of                 
quantization
of the PBHT (\ref{hyd}). Its decision may be find
by means of the change of
variables $u=u(v)$ such that in the new variables $v$
bracket (\ref{hyd}) is constant. After canonical
quantizations of the constant bracket
we can come back to the
old variables $u$. But we run up against
the problem of ordering. Thus
we need the simplest as it is possible change of the variables.
The linear change is not suitable.
Since according to \cite{DN1} in the
case of nondegenerate  algebras
(\ref{alg}),(\ref{myalg1})
the metric $g^{ij}=C^{ij}_{k}u^{k}$
must have zero curvature,
we appeal to changes $u(v)$, which are now nonlinear,
where metric in the new coordinates $(v^{1}, \ldots ,v^{N})$
is constant
    \[g^{ij}(u(v))=g^{\alpha \beta}_{0}(\partial
      u^{i}/ \partial v^{\alpha})(\partial
     u^{i}/ \partial v^{\beta}), \ \ \ \ g^{\alpha \beta}_{0}=const.  \]
We consider the purely quadratic changes
\begin{equation}
   u^{i}=\frac{1}{2}F^{i}_{\alpha \beta}v^{\alpha}v^{\beta}.  \label{q}
\end{equation}
Then
for a change (\ref{q}) to reduce the nondegenerate
metric of zero curvature $g^{ij}=C^{ij}_{k}u^{k}$
( from PBHT ) to constant form  it is necessary and
sufficient that the following conditions hold
\begin{itemize}
    \item  $b^{ij}_{k}=b^{ji}_{k}$;
    \item  $F$ and $g^{ij}_{0}$ determine
           a Frobenius representation of the
           algebra (\ref{myalg1}), where
           the $ F^{i}_{\alpha \gamma} $ give
           a representation of the basis $e^{i}$ of
           the algebra in the form of linear
           operator in $v$-space which are selfadjoint
           in this inner product, so that
   \[e^{i}\longrightarrow (F^{i})^{\alpha}_{\beta}=g^{\alpha}_{\gamma}
    F^{i}_{\gamma \beta}, \ \ \ \ F^{i}F^{j}=C^{ij}_{k}F^{k}/2 \]
    \[ det (F^{i}_{\alpha \beta}v^{\beta} \not\equiv 0).   \]
\end{itemize}
Thus if the algebra $\bf B$ (\ref{myalg1}) is commutative
and nondegenerate (in this case it has the unit) by the
quadratic changes we obtained constant bracket.
After the linear change of variables from $v$ to $(\phi)^{'}$
we have
\[ \{( \phi ^{i}(x))^{'}, (\phi ^{j}(y))^{'} \}
 = \delta ^{ij} \delta{'}(x-y).     \]
Dirac
quantization lead to the theory of free fields  in 2D quantum
field theory, and the Fourier components of $(\phi)^{'}$
form  the following famous algebra($i=1, \ldots ,N,\ s \in \mbox{\bf Z}$):
\[ [a^{i}(s),a^{j}(k)]=k \delta ^{ij} \delta _{s+k,0} .  \]
  After choice the
ordering procedure for $a^{i}(s)$ we obtain
for (\ref{q}) not algebra (\ref{W}),
but the algebra (\ref{vir}), and the ordering
procedures are in the one to one correspondence with
central extensions of (\ref{W}).
Thus algebra (\ref{vir})
is the quantization of PBHT (\ref{hyd})
in the case when  algebra (\ref{myalg1}) is
symmetric and nondegenerate.
                                                   \\

{\bf Remark} \ \ \ It will be very interesting                        
investigate the algebras
(\ref{myalg1}) and PBHT  with another
type of the change of
variables $u=u(v)$ such that in the
new variables $v$
bracket (\ref{hyd}) is constant .
Some examples of such noncommutative
algebras (for the Poisson brackets
of one-dimensional hydrodynamics) were
investigated in \cite{BalNov}.
                                                    \\

 We are very grateful to Professor S P Novikov
 for  numerous discussion and suggestion.

\pagebreak

\end{document}